\documentclass{article}
\usepackage{emulateapj,amsfonts,psfig}

\newcommand{\be}{\begin{equation}}
\newcommand{\en}{\end{equation}}
\def\ltsima{$\; \buildrel < \over \sim \;$}
\def\lsim{\lower.5ex\hbox{\ltsima}}
\def\gtsima{$\; \buildrel > \over \sim \;$}\def\gsim{\lower.5ex\hbox{\gtsima}}
\def\deg {^\circ}
\def\mdot {\dot M}

\def\ergs  {\rm \ erg \, s^{-1}}

\def\gs   {\rm \ g  \, s^{-1}}

\def\cmdue {\rm \ cm^{-2}}

\def\msole {~M_{\odot}}

\begin{document}

\title{The transient X--ray pulsar 4U~0115+63 from quiescence to outburst
through the centrifugal transition}

\author{S. Campana\altaffilmark{1}, F. Gastaldello\altaffilmark{2,1},
L. Stella\altaffilmark{3,4}, G.L. Israel\altaffilmark{3,4}, 
M. Colpi\altaffilmark{2}, F. Pizzolato\altaffilmark{2}, 
M. Orlandini\altaffilmark{5}, D. Dal Fiume\altaffilmark{5}}

\altaffiltext{1}{Osservatorio astronomico di Brera, Via Bianchi 46, 
I-23807 Merate (LC), Italy}
\altaffiltext{2}{Dipartimento di Fisica G. Occhialini, Universit\`a di
Milano Bicocca, Piazza della Scienza 3, I-20126 Milano, Italy}
\altaffiltext{3}{Osservatorio astronomico di Monteporzio Catone, 
Via Frascati 33, I-00040 Monteporzio Catone (Roma), Italy}
\altaffiltext{4}{Affiliated to I.C.R.A.}
\altaffiltext{5}{TeSRE, C.N.R., Via Gobetti 101, I--40129 Bologna, Italy}

\authoremail{campana@merate.mi.astro.it}

\begin{abstract}
We report on a BeppoSAX observation of the transient X--ray pulsar 4U 0115+63 
close to periastron. This led to the discovery of a dramatic luminosity variation 
from $\sim 2\times 10^{34}\ergs$ to $\sim 5\times 10^{36}\ergs$ (factor $\gsim 250$) 
in less than 15 hr. The variation was accompanied by 
only minor (if any) changes in the emitted spectrum and pulse fraction. 
On the contrary an observation near apastron detected the source in a nearly 
constant state at a level of $\sim 2\times 10^{33}\ergs$. 
Direct accretion onto the neutron star surface encounters major 
difficulties in explaining the source variability properties. 
When the different regimes expected for a rotating magnetic neutron star subject 
to a variable inflow of matter from its companion are taken into consideration, 
the results of BeppoSAX observations of 4U 0115+63 can be explained 
naturally. In particular close to apastron, the regime of centrifugal inhibition 
of accretion applies, whereas the dramatic source flux variability observed close 
to periastron is readily interpreted as the transition regime between direct 
neutron star accretion and the propeller regime. In this centrifugal transition 
regime small variations of the mass inflow rate give rise to very large 
luminosity variations. We present a simple model for this transition, which we 
successfully 
apply to the X--ray flux and pulse fraction variations measured by BeppoSAX. 
\end{abstract}


\section{Introduction}

Accreting collapsed stars in transient X--ray binaries are subject to 
very large variations of mass inflow rate and
provide a laboratory to test the physics of accretion 
over a range of different regimes that is unaccessible to persistent sources 
(see e.g. Parmar et al. 1989; Campana et al. 1998).
The highly magnetic ($B\sim 10^{12}$ G), spinning neutron stars that are hosted 
in Hard X--ray Transients (HXRTs) have long been suspected to be  
in the so-called propeller regime when quiescent. In this regime 
accretion onto the neutron star surface is inhibited by the centrifugal action 
of the rotating magnetosphere (Illarionov \& Sunyaev 1975). On the other hand 
there is little doubt that accretion onto the neutron star surface takes place 
in these systems while in outburst, as they display very similar properties 
to those of persistent X--ray pulsars. 

Relatively little is known on the low luminosity states of HXRTs.  
Observations of their quiescent state are still sparse (Campana 1996; Negueruela 
et al. 2000) and there are only hints that the transition from the lowest 
outburst luminosities to quiescence occurs in a sudden fashion (see e.g. the case 
of V 0332+53, Stella et al. 1986). According to models of accretion 
onto rotating magnetic neutron stars (e.g. Illarionov \& Sunyaev 1975), the transition 
from the accretion to the propeller regime (or vice versa) should take place over a 
very limited range of mass inflow rates and give rise to luminosity variation in 
the order $\sim 1000$ for neutron star spinning at a few second period. 
The observation of a paroxysmal luminosity variation when a HXRT emerges out 
of quiescence or fades away at the end of an outburst therefore holds a great 
potential as a diagnostic of the different regimes experienced by a rotating 
magnetic neutron star.

In this paper we report on BeppoSAX observations of the transient X--ray 
pulsar 4U~0115+63, which reveal the quiescent state of the source 
and, more crucially, provide the first convincing evidence for the centrifugal 
transition regime of any X--ray transient. 

\begin{table*}[!htb]
{\footnotesize
\label{spectra}
\caption{Exponential cut-off power law fits to spectra from different intervals 
from the August 3--4, 1999 observation.}
\begin{tabular}{cccccccc}
\hline
Interval & Time      & $N_H$                    & Power law              &$E_{\rm cutoff}$     & $E_{\rm fold}$      & 0.1--200 keV Flux$^\S$            &$\rm{\chi_{red}^2}$\\
         & (hr)      &($10^{22}$ $\rm{cm^{-2}}$)& $\Gamma$               &(keV)                &(keV)                & (erg s$^{-1}$ cm$^{-2}$)          &   (d.o.f.)        \\ 
\hline
$1^+$    &\ 7.5--14.9& $2.30^{+1.08}_{-0.97}$   & $1.58^{+0.40}_{-0.40}$ & 12.0 (fixed)          & 8.2 (fixed)         & $1.1\times 10^{-11}$& 1.52 (24)  \\
2        & 15.4--17.2& $1.83^{+0.72}_{-0.57}$   & $0.82^{+0.18}_{-0.16}$ &$14.3^{+2.5}_{-4.8}$ &$\ 6.3^{+5.2}_{-2.2}$& $1.5\times 10^{-10}$& 1.13 (28)   \\
3        & 17.5--19.4& $1.80^{+0.80}_{-0.64}$   & $1.00^{+0.23}_{-0.20}$ &$11.7^{+12.0}_{-3.9}$&$13.2^{+10.5}_{-8.3}$& $1.5\times 10^{-10}$& 1.07 (33)  \\
4        & 20.5--21.5& $1.72^{+0.40}_{-0.37}$   & $0.90^{+0.13}_{-0.13}$ &$\ 9.7^{+3.7}_{-1.6}$&$\ 8.5^{+2.3}_{-3.3}$& $2.8\times 10^{-10}$& 1.04 (43)   \\ 
5        & 22.0--22.5& $1.41^{+0.35}_{-0.30}$   & $0.89^{+0.12}_{-0.11}$ &$10.3^{+4.0}_{-2.0}$ &$\ 8.0^{+2.8}_{-3.6}$& $5.0\times 10^{-10}$& 0.97 (40)   \\
$6^{\dag}$&22.5--22.7& $1.74$ (fixed)           & $1.10^{+0.13}_{-0.13}$ &$12.5^{+4.9}_{-6.3}$ &$\ 6.0^{+6.6}_{-2.9}$& $5.7\times 10^{-10}$& 0.78 (41)   \\
$7^{\dag}$&22.7--23.3& $1.74$ (fixed)           & $0.95^{+0.13}_{-0.13}$ &$13.2^{+4.4}_{-2.2}$ &$\ 5.5^{+3.7}_{-2.1}$& $7.3\times 10^{-10}$& 1.48 (36)   \\
8        & 23.3--23.9& $1.83^{+0.46}_{-0.36}$   & $0.94^{+0.17}_{-0.14}$ &$\ 9.4^{+4.2}_{-1.8}$&$11.2^{+5.1}_{-4.4}$ & $1.1\times 10^{-9}$ & 1.02 (33)    \\
9        & 23.9--24.2& $1.48^{+0.45}_{-0.39}$   & $0.85^{+0.15}_{-0.15}$ &$11.7^{+2.7}_{-4.2}$ &$\ 7.8^{+2.4}_{-1.8}$& $8.9\times 10^{-10}$& 0.98 (44)   \\
$10^{\dag}$&24.2--24.4&$1.74$ (fixed)           & $0.93^{+0.11}_{-0.11}$ &$\ 9.1^{+3.4}_{-2.3}$&$\ 9.7^{+2.4}_{-2.8}$& $9.0\times 10^{-10}$& 0.88 (44)   \\ 
11       & 24.5--25.7& $3.45^{+1.00}_{-0.84}$   & $1.30^{+0.21}_{-0.21}$ &$15.0^{+4.0}_{-7.0}$ &$\ 9.4^{+4.2}_{-3.1}$& $7.3\times 10^{-10}$& 1.13 (36)   \\
\hline
Mean value  & & $1.74\pm0.18$          & $0.95\pm0.05$          & $12.0\pm1.3$        & $8.2\pm1.1$        &                     &             \\ 
$\chi^2_{\rm red}(\rm cons)$ & & 1.7   &  2.1                   & 0.8                 & 0.6                &                     &             \\  
\hline
\end{tabular}

\noindent 
Time is measured in hours from the start of MJD 51393.

\noindent
$^\S$ Fluxes are unabsorbed. Errors are at $90\%$ confidence level (c.l.) for one 
parameter of interest (i.e. $\Delta\chi^2=2.71$).

\noindent
$^+$ Due to poor statistics, we fixed the cut-off and folding energies to the
mean value of the entire observation. In the case of a simple power law fit,  
a column density of 
$(2.2^{+1.1}_{-0.9})\times 10^{22}\cmdue$ and a photon index of 
$\Gamma=1.54^{+0.42}_{-0.33}$ ($\chi_{\rm red}^2=1.44$) were obtained.

\noindent
$^{\dag}$ Only MECS and PDS data available. The column density has been fixed
to the mean value since without the LECS data it is difficult to constrain
its value.}
\end{table*}

\section{4U 0115+63}

Like most HXRTs, 4U~0115+63 hosts a magnetic neutron star orbiting a Be 
star in a moderately eccentric orbit ($e\simeq0.3$). 
The spin and orbital periods are $P_0=3.62$~s and $P_{\rm orb}\simeq
24.3$~d (Cominsky et al. 1978; Rappaport et al. 1979). 
The system has been extensively studied during its frequent outbursts
($\sim 20$ have been detected so far), which occasionally reach a peak luminosity of  
$\sim 10^{38}\ergs$ and usually last about a month (Bildsten et al. 1997; Campana 1996). 
The neutron star magnetic field is measured with good accuracy, $B_0=1.3
\times 10^{12}\,[(1+z)/1.2]$ G, through the cyclotron features 
detected in its X--ray spectrum (Santangelo et al. 1999 and references therein).
Here $(1+z)=1+G\,M/(c^2\,R)\simeq 1.2$ (we scale the neutron star mass and
radius as $M=1.4\,M_{1.4}\msole$ and $R=10\,R_6$ km, respectively) is the 
gravitational redshift at the neutron star surface. 
The source distance was determined to be $(8\pm1)\,d_8$ kpc, such that the X--ray flux 
to luminosity conversion is correspondingly accurate (Negueruela \& Okazaki 2001).

Before the present study, 4U 0115+63 was not detected in its quiescent state 
to a limit of $\sim 6\times 10^{33}\ergs$ (0.4--6 keV; Campana 1996). At the end 
of the 1990 outburst a quick decrease of the luminosity was observed below a level 
of $\sim 5\times 10^{36}\ergs$, possibly indicating the transition to the 
onset of the centrifugal barrier (Tamura et al. 1992). 

\section{BeppoSAX Observations} 

\begin{figure*}[!htb]
\psfig{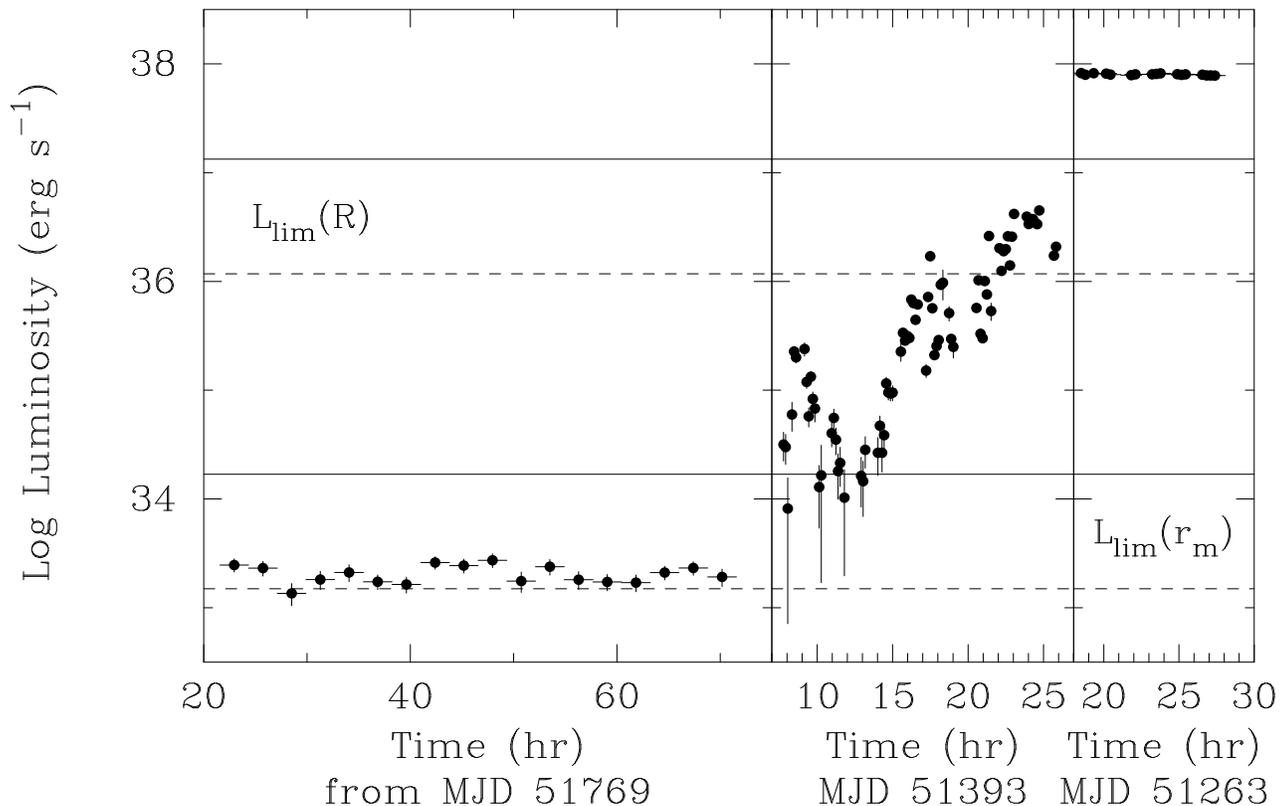}
\caption{ 
Light curve of the BeppoSAX MECS observations of 4U 0115+63 taken in quiescence 
(left), during the transition (middle) and in outburst (right). Background
subtracted light curves were converted into luminosities using the conversion 
factor derived from spectral fits (see text). Time bins are 10,000 s, 500 s and 
5,000 s for the three observations, respectively. 
The time axis values (in hr) correspond to the start of the 
three observations discussed in the text (MJD 51769, 51393 and  51263, 
respectively). Solid lines represent 
the luminosity corresponding to the onset of the centrifugal barrier 
[$L_{\rm lim}(R)$], and the luminosity at which it closes completely
[$L_{\rm lim}(r_{\rm m})$] for $\xi=1$. Dashed lines give
the same luminosities for $\xi=0.5$.
}
\label{fig1}
\end{figure*}

A 35 ks observation of 4U~0115+63 was carried out with the Italian/Dutch 
satellite BeppoSAX (Boella et al. 1998) on 1999 August 3 06:37:04 UT -- August
4 01:41:32 UT, covering an orbital phase range (0.94--0.97) close to periastron. 
The source count rate varied by a very large amount (from $\gsim 0.02$ to $\sim 5$ 
c s$^{-1}$) showing a steep increase by a factor of $\gsim 250$ 
in $\sim 15$~hr (see Fig. \ref{fig1}; note that the Y-axis is
logarithmic). Superposed 
to this trend, variations of up to a factor of $\sim 20$ in less than one hour were 
clearly present.

Inspection of the RossiXTE ASM light curve of 4U 0115+63 shows that the dramatic 
flux increase revealed by the BeppoSAX observation was not followed by a source 
outburst over the following days; the source remained instead below a daily average 
of $\sim 2\times 10^{36}\ergs$.

In consideration of the extreme variability, we divided the 
entire observation in different time intervals (for spectral analysis) as well 
as intensity intervals (for temporal analysis\footnote{Pulsations were detected 
also at the lowest intensity interval, even though the source spectrum could not 
be meaningfully characterized; this is why time intervals were used instead 
in the spectral analysis.}), in order to characterize 
the source properties at different luminosity levels.
Energy spectra from the BeppoSAX LECS (0.1--4 keV), MECS (1.8--10 keV) and PDS 
(15--200 keV) were accumulated in each of these intervals.
In all cases the source spectrum from the three instruments was well fit by a 
model consisting of an absorbed power law with high-energy cut-off plus 
absorption, with photon index of $\Gamma \sim 1.0$, cut-off energy 
$E_{\rm cut} \sim 12$ keV, $e-$folding energy $E_{\rm fold}\sim 8$ keV. 
The column density amounts to $N_H\sim 1.7\times 10^{22}\cmdue$ and it is 
consistent with the galactic value. This model fits also nicely the 
spectrum of the entire observation and is compatible also with 
the source spectrum measured during outburst (see e.g. Nagase 1989). 
The best fit parameters are reported in Table 1 for each interval 
(see also Gastaldello et al. 2001, in preparation). 
We note that all spectral parameters are nearly constant throughout the 
observation (see Table 1). Variations are observed only in the column density 
($4\%$ probability of getting a higher $\chi^2$ by chance from a constant distribution)
and in the power law index (probability of $1\%$).
The main contribution to the variations of the power law index comes from the
first interval. If we exclude this value, we obtain a probability of $10\%$. 
Marginal evidence was found for the second cyclotron harmonic 
at a centroid energy of $22.5\pm2.5$ keV in the spectrum 
from the entire observation 
($95\%$ significance; Santangelo et al. 1999 measured $24.16\pm0.07$ keV
during outburst). Note that the first harmonic 
falls just in the gap between the MECS and the PDS spectra.

\begin{figure*}[!htb]
\vskip -9truecm
\psfig{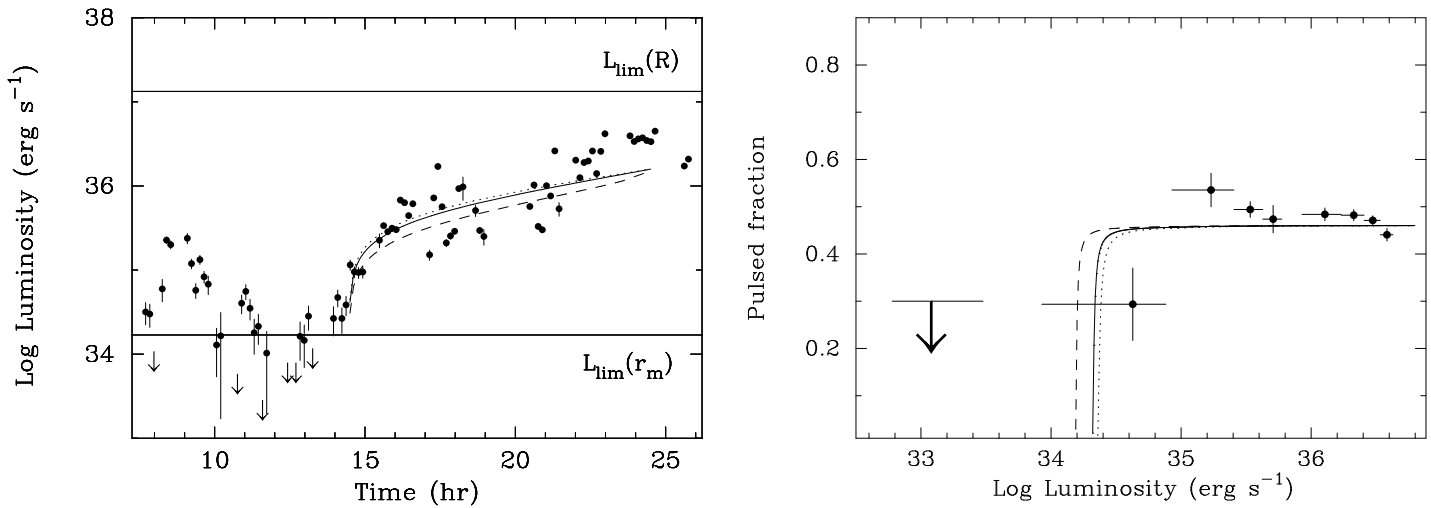}
\caption{Left panel: 
Light curve of the 0.1--200 keV unabsorbed luminosity of 4U 0115+63
versus time during the August 3--4, 1999 BeppoSAX observation. Arrows indicate 
500 s intervals for which only upper limits ($3\,\sigma$) can be set. 
The solid lines are the same as in Fig. \ref{fig1}. The other curves show different 
applications of the stellar wind and centrifugal transition model 
presented in Section 4. The dashed, solid 
and dotted lines refer to angles $\chi=28^{\rm o},\, 35^{\rm o},\, 80^{\rm o}$,
respectively. These are the minimum angle compatible with the assumed
mass inflow rate variation ($28^{\rm o}$), the maximum value (we take $80^{\rm
o}$ since $90^{\rm o}$ provides the same result but more computational
problems) and an angle providing a curve in between ($35^{\rm o}$).
In the fit, $\xi=1$ and the Be star equatorial disk model described 
in Section 4 have been assumed. 
Right panel: Pulsed fraction versus 0.1--200 keV unabsorbed 
luminosity during the BeppoSAX observations. Pulsed fractions were 
obtained by fitting the folded light curve with a sinusoidal signal. Error 
bars are a $1\,\sigma$. The curves represent the pulsed 
fraction $k$ from the model presented in Section 4. No free parameter is involved 
except for the normalization $k_{\rm mag}$ ($=0.5$). The dashed, solid 
and dotted lines refer to angles $\chi=28^{\rm o},\, 35^{\rm o},\, 80^{\rm o}$,
respectively. The upper limit ($90\%$) refers to the BeppoSAX observation of 
Aug 13--16, 2000.}
\label{fig2}
\end{figure*}

In this letter we estimate the source bolometric luminosities by using 
the 0.1--200 keV unabsorbed luminosities as derived 
from the best fit parameters of the spectrum from the entire observation 
and the MECS count rates in the 1.8--10 keV band. 
The corresponding conversion factor
was $1\, {\rm c\, s^{-1}} = 8.5\times 10^{35}\ergs$. 
The luminosity inferred in this way varied
between $2\times 10^{34}$ and $5\times 10^{36}\,d_8^2\ergs$ (see Fig. \ref{fig2}).

Pulsations at the $\sim 3.62$ s neutron 
star spin period were detected in all intensity intervals.
The pulsed fraction (semi-amplitude of modulation divided by the mean source
count rate) in the 1.8--10 keV energy band was determined for each interval
after folding the data at the best period. 
Values ranged between $\sim 29\%$ and $\sim 54\%$ 
with the lowest value corresponding to the lowest intensity interval
(see Fig. \ref{fig2}). 

A further 86 ks BeppoSAX observation was carried out on 2000 August 13 21:33:06 UT 
-- August 16 00:22:36 UT around apastron (orbital phase of 0.42--0.51).
The source was very weak and remained at a virtually constant level of 
$\sim 2\times 10^{-3}$ c s$^{-1}$ in the MECS; it was not detected in 
the LECS ($<2\times 10^{-3}$ c s$^{-1}$). 
The spectrum was softer than in the August 1999 observation.
Fitting a power-law model to the MECS data and fixing the column density to the
best fit value of the periastron observation yielded a photon index of 
$\Gamma=2.6^{+1.0}_{-0.8}$ (at $68\%$ c.l., $\chi_{\rm red}^2=0.2$ for 2 d.o.f.).
The uncertainties in the spectral parameters 
translate also into a fairly large uncertainty in the inferred 0.1--200 keV 
unabsorbed luminosity, which is $(0.6-3)\times 10^{33}\ergs$. 
A pure blackbody model gave also a reasonable fit for a temperature
of $k\,T=0.7^{+0.3}_{-0.2}$ keV ($68\%$ c.l., $\chi^2_{\rm red}=0.5$).

The $\sim 3.62$ s pulsations were not detected; a $3\,\sigma$ upper 
limit of $\sim 30\%$ on the pulsed fraction was derived 
(see also Campana et al. 2001 in preparation).

In order to facilitate the comparison with the source properties while in 
outburst we also analyzed the data from the 48 ks BeppoSAX observation 
that took place on 1999 March 26, during the outburst decay.
This observation covered an orbital phase interval of 0.61--0.65. 
The source light curve is also shown in Fig. \ref{fig1} for comparison; the average 
0.1--200 keV luminosity was $\sim 8\times 10^{37}\ergs$.  
The source spectrum could be described by a power law with a high energy cut-off
with $\Gamma \sim 0.8\pm0.1$, $E_{\rm cut} \sim 9.4\pm0.5$ keV, 
$E_{\rm fold}\sim 16\pm 2$ keV and $N_H 
\sim (1.5\pm 0.1) \times 10^{22}$ cm$^{-2}$. To obtain an acceptable fit also 
an iron line and a cyclotron line feature had to be included.

\

\section{Evidence for the transition from the propeller to the accretion regime}

The factor of $\gsim 250$ luminosity variation of 4U~0115+63 
during the August 1999 close to periastron observation, when the source luminosity 
ranged between $2\times 10^{34}$ and $5\times 10^{36}\ergs$ in $\sim 15$~hr 
is the most extreme ever seen in a HXRT.
Previously the steepest flux increases detected from this source (and other 
HXRTs) were those associated to the rise to an outburst peak, involving 
variations of up to a factor of $\sim 3-4$ on a comparable time scale. 
On the contrary during the August 2000 observation close to apastron  
the source luminosity was low and nearly constant around 
a level of $2\times 10^{33}\ergs$.  

One possibility to explain the extreme variability in the 
August 1999 observation is that the source was 
subject to factor of $\gsim 250$ variations in the mass inflow rate, which in 
turn gave rise to a comparable variation in accretion luminosity (under the 
hypothesis that accretion onto the neutron star surface took place unimpeded 
also for luminosities as low as $\sim 10^{34}\ergs$). This possibility faces serious 
problems. Firstly, models of Be star disks/winds predict a neutron star 
mass capture rate variation of a factor of $\sim 5$ at the most for a binary 
system such as 4U~0115+63 over an orbital phase interval of 0.94--0.97
(Raguzova \& Lipunov 1998). 
Moreover, within the direct accretion scenario (required by interpretation above),  
the issue as to why extreme variability manifests itself over a range 
of intermediate luminosities (between $2\times 10^{34}$ and $5\times 10^{36}\ergs$), 
while the source variability is much less pronounced for both higher and 
lower luminosities, remains unaddressed. 

The behaviour of 4U 0115+63 has a much more natural explanation in terms 
of the different regimes that are expected for a rotating magnetic neutron star
subject to a variable mass capture rate. 
The inflow of matter towards an accreting magnetic star is dominated by 
gravitational forces and possibly mediated by an accretion disk 
down to the magnetospheric boundary at $r_{\rm m}$.
 From this radius matter can proceed inwards to the star surface
{\it only if} the centrifugal force due to magnetospheric drag is
weaker than gravity. 
This translates into the condition that the radius at which a test
particle in Keplerian circular orbit corotates with the central object
[the so-called corotation radius, $r_{\rm cor}= (G\,M\,P^{2}/4\,
\pi^2)^{1/3}$, with $P$ the neutron star spin period and $G$ the gravitational 
constant] is larger than $r_{\rm m}$ (Illarionov \& Sunyaev 1975). 
In this direct accretion regime, the matter to radiation conversion efficiency 
is high $L(R) = G\,M\,\mdot/R$ and gives rise to luminous X--ray sources
(here $\mdot$ is the mass inflow rate).
If $r_{\rm m} >r_{\rm cor}$ the drag by the rotating 
magnetosphere is super-Keplerian, such that the centrifugal 
force exceeds gravity and accretion is inhibited. 
However, as the inflowing matter reaches $r_{\rm m}$, an accretion 
luminosity of $L(r_{\rm m}) \simeq G\,M\,\mdot/(2\,r_{\rm m})$ must be released
(the factor of 1/2 comes from assumption that the flow down to
$r_{\rm m}$ is mediated by a disk). 
$L(r_{\rm m})$ is expected to scale approximately as $\sim \mdot^{9/7}$. 
Therefore, above and below the transition to the propeller regime a smooth,
close to linear, relationship between the accretion luminosity and $\mdot$ is
expected.

An important prediction is that the accretion luminosity across 
the transition from direct accretion to the propeller regime
(corresponding to $r_{\rm m} \simeq r_{\rm cor}$) or vice versa
should be characterized by a sudden luminosity jump
\begin{eqnarray}
L_{\rm lim}(R)/L_{\rm lim}(r_{\rm m}) &\simeq& 2\,r_{\rm cor}/R \\ \nonumber
&&\simeq 2\,\Bigl( {{G\,M\,P^2}\over {4\,\pi^2\,R^3}}\Bigr)^{1/3}
\end{eqnarray}
This is a factor of $\sim 800$ in the case of 4U 0115+63 
(Corbet 1996; Campana \& Stella 2000).
In practice the transition separating the two regimes 
is expected to take place over a finite, though small, range of mass inflow 
rates around $\mdot_{\rm lim}$, such that a very steep dependence of the 
accretion luminosity on $\mdot$ ensues temporarily. 

Since the magnetospheric radius changes in response to variations of the mass inflow 
rate, the absolute luminosity at which the centrifugal barrier is expected to close 
can be determined based on models of the interaction between the inflowing 
matter and the rotating magnetosphere. In a simple spherical accretion 
approximation $r_{\rm m}$ is expected to scale as $\mdot^{-2/7}$ (Davidson \& 
Ostriker 1973). 
A similar dependence is obtained in the detailed models of the disk/magnetosphere 
that were developed by (e.g.) Ghosh \& Lamb (1979) and Wang (1996) in the 
regime in which the disk is dominated by gas pressure (the one relevant to the 
case of 4U 0115+63). 
In order to account for the predictions of different models, 
we adopt the scaling above and introduce a correction factor 
$\xi$, the ratio of the magnetospheric radius determined on the basis of a 
given model to that of simple spherical accretion. We obtain  
\be
% 1.33
L_{\rm lim}(R) \simeq 1\times 10^{37}\,\xi^{7/2}\,B_0^2\,P_0^{-7/3}\,M_{1.4}^{-2/3}
\,R_6^{5}\ergs 
\en
where $\xi=1$ for spherical accretion (by definition), $\xi=0.5$ in the model 
by Ghosh \& Lamb (1979) and $\xi\sim 1$ in the model by Wang (1996).We take
this range as representative of the accuracy with which $r_{\rm m}$ can be 
predicted by current models ($M_{1.4}$ and $R_6$ are
the neutron star mass and radius in units of 1.4 solar masses and $10^6$ cm, 
respectively. See e.g. Campana et al. 1998).
Since the source distance (and, therefore, luminosity) and neutron star magnetic 
field are fairly accurately measured in the case of 4U~0115+63, we 
obtain $L_{\rm lim}(R) \simeq \xi^{7/2}\, 10^{37}\ergs$; correspondingly 
$L_{\rm lim}(r_{\rm m}) \simeq \xi^{7/2}\, 2\times 10^{34}\ergs$. 
%\be
%%3.37
%L_{\rm lim}(r_{\rm m}) \simeq 3\times 10^{34}\,\xi^{7/2}\,B_0^2\,P_0^{-3}
%\,M_{1.4}^{-1}\,R_6^{6}\ergs 
%\en
The corresponding lines for $\xi=1$ (continuous) and $\xi=0.5$ (dashed) are 
shown in Fig. \ref{fig1}. 
It is immediately apparent that the variations of the August 
1999 observation fall just in the luminosity range of the transition 
between the propeller and the direct accretion regime, where very large 
luminosity variations are expected in response to modest changes 
of the mass inflow rate. 
On the contrary the quiescent state luminosity of the August 2000 observation,
during which the source flux was approximately constant, lies in the propeller 
regime, while the luminosity during the outburst observation on March 1999 is 
well in the range of the direct accretion regime. 

\section{A Simple Model for the Centrifugal Transition Regime}

In this section we develop a simple model for the centrifugal transition regime 
and compare its predictions to the results from the 
August 1999 observation of 4U 0115+63. 
In general, we express the source accretion luminosity in the transition  
regime as the sum of two contributions: (a) the luminosity of the disk 
extending down to the magnetospheric boundary, $L_{\rm disk}$;  
we assume that mass flows through the disk at rate $\mdot$, which is 
equal to the rate at which mass is captured at the neutron star accretion 
radius; (b) the luminosity released within the magnetosphere
$L_{\rm mag}$ by the fraction $f$ of the mass inflow 
rate that effectively accretes onto the neutron star surface. We have 
\begin{eqnarray}
L &=& L_{\rm disk} + L_{\rm mag} = L(r_{\rm m}) + f(\,L(R) - L(r_{\rm m})) \\ 
\nonumber &=&
G\,M\,\mdot\, [1/2\,r_{\rm m} + f\,(1/R - 1/2\,r_{\rm m})] 
\end{eqnarray}
This approximation the direct accretion regime corresponds to $f=1$ and the 
propeller regime to $f=0$.

In the case of 4U 0115+63, being $2\,r_{\rm cor}\sim 800\,R$,  
$f\,L(R) > (1-f)\,L(r_{\rm m})$ for $f \gsim 10^{-3}$, i.e. 
the luminosity produced by matter accreting onto the neutron star surface
is dominant over most of the centrifugal transition. 
This has two important implications. Firstly, 
being dominated by the release of energy within the magnetosphere,
the emitted X--ray spectrum should 
be similar to that observed in the direct accretion regime (i.e. source
outbursts) for luminosities $\leq 10^{37}\ergs$ and remain nearly 
unchanged across most of the centrifugal transition. We note also that according 
to models of accretion columns onto magnetic neutron stars, the spectrum   
is virtually insensitive to accretion rate variations as long as the 
optical depths remains $<1$ (Nagel 1981). 
Only for luminosities of $\lsim 10^{34}\ergs$ 
(corresponding to $f \lsim  10^{-3}$), when $L_{\rm disk}$
becomes non-negligible, substantial spectral changes are to be expected. 
The apparent (relative) stability of the BeppoSAX spectra during the 
August 1999 observation are consistent with this expectation. 
Secondly, as the transition regime involves only a relatively small variation 
of $r_{\rm m}$ the geometry of accretion close to the polar caps 
should show only minor changes
(Wang \& Welter 1981; Parmar et al. 1989). 
Therefore the pulse fraction is expected to remain essentially unaltered 
as long as $L_{\rm mag}$ dominates. 
On the other hand, $L_{\rm disk}$ is expected to be unpulsed and 
the source pulsed fraction should decrease close to bottom of the transition 
regime. This is also consistent with the results from the August 1999 
observation of 4U 0115+63 (see Table 1 and Fig. \ref{fig2}).

\begin{figure*}[!htb]
\vskip -5truecm
\psfig{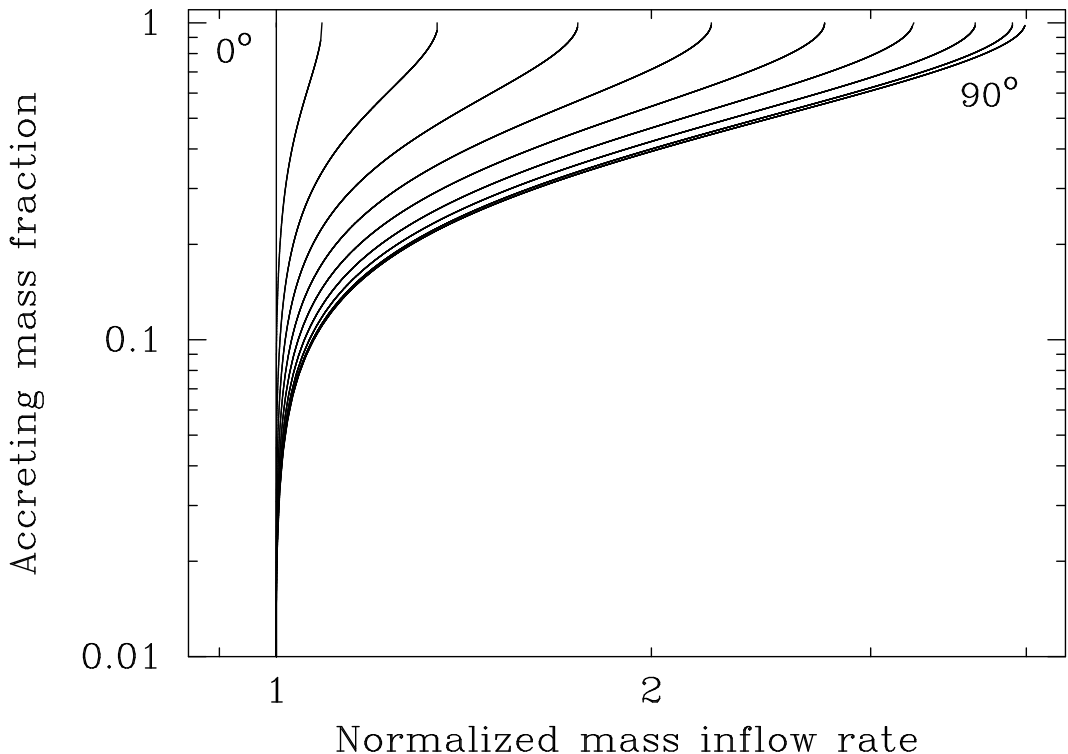}
\caption{Fraction of accreting mass $f$ as a function of the mass inflow rate
$\mdot$ normalized to the maximum mass inflow rate in the propeller regime,
on the basis of the model discussed in Section 5.
The different curves refer to angles $\chi$ between $0\deg$ (left) 
to $90\deg$ (right) in step of $10\deg$. 
}
\label{fig3}
\end{figure*}

We explored a simple model to determine $f$ and the source behaviour across the 
centrifugal transition: we considered a magnetic dipole field the axis 
of which is tilted relative to the neutron star rotation axis by an angle $\chi$.
We determined the azimuthal ($\phi$-)dependence of the magnetospheric boundary 
in the disk plane by equating the ram pressure of radially free-falling matter 
with the local magnetic pressure. 
The magnetospheric boundary takes an elongated shape given by 
\begin{eqnarray}
r_{\rm m}(\mdot,\phi)&=&\xi\,B^{4/7}\,R^{12/7}\,\\ \nonumber
&&[1+3\sin^2\chi\sin^2\phi]^{2/7}\,(2\,G\,M\,\mdot^2)^{-1/7} 
\end{eqnarray}
(Jetzer, Str\"assle \& Straumann 1998).
The minimum radius $r_{\rm m}(\mdot,0)$ corresponds also 
to $\chi=0$, the approximation 
usually adopted in models of the disk-magnetosphere interaction.
The maximum radius $r_{\rm m}(\mdot,\pi/2)$ is only a factor of 
$[1+3\sin^2 \chi]^{2/7}\leq 1.49 $ larger. 
If $r_{\rm cor} < r_{\rm m}(\mdot,0)$ the magnetospheric boundary is larger
than the corotation radius for any $\phi$ and the propeller regime applies 
(i.e. $f=0$). If on the contrary  $r_{\rm cor} > r_{\rm m}(\mdot,\pi/2)$
every point on the boundary is within the corotation radius and the 
standard accretion regime applies ($f=1$). In the intermediate regime,  
$r_{\rm m}(\mdot,0) < r_{\rm cor} <  r_{\rm m}(\mdot,\pi/2)$, only the fraction 
$f = \Delta\phi/2\pi$ of the magnetospheric boundary for which 
$r_{\rm cor} > r_{\rm m}(\mdot,\phi)$ leads to accretion onto the neutron star 
surface and the transition regime applies. The accretion luminosity versus
mass inflow rate curves have been calculated by using this model with $\xi=1$.
Different curves correspond to different values of the angle
$\chi$, the sharpest transitions occurring for the lowest values 
of $\chi$, as expected (see Fig. \ref{fig3}).

We adopted the model above also to describe the evolution of the source pulse 
fraction versus luminosity (see Fig. \ref{fig2}). If $L_{\rm disk}$ is unpulsed, the
observed pulsed fraction $k$ can be expressed as 
$k \simeq k_{\rm mag}\,f/(R/2\,r_{\rm m} + f)$,   
where $k_{\rm mag}$, the pulsed fraction of $L_{\rm mag}$ alone, can 
effectively be regarded as a free parameter. $k$ depends on $\xi$ 
(which sets the onset luminosity of the transition regime) and 
$\chi$ (which determines the range of the mass inflow rates over which 
the transition regime applies). 
Here too the dependence on $\chi$ is such that for low values the transition is 
sharp, whereas for large values the transition occurs over a mass inflow rate
variation by $\sim 4$ (see Fig. \ref{fig3}).

Modelling the time evolution of the source luminosity during the August 1999
observation would require knowing the time evolution of $\mdot$. 
We approximated $\mdot(t)$ with the mass capture rate 
by the neutron star 
in its orbital motion within the equatorial disk of the Be star 
companion over the phase interval 0.94--0.97. 
We assumed a 10 km s$^{-1}$ nearly Keplerian disk-outflow velocity and a 
density profile $\propto r^{-3}$ (as in Raguzova \& Lipunov 1998).
This model predicts a $\mdot$ variation of $\lsim 3$ over the phase interval 
covered by the August 1999 observation. We remark that this is among 
the largest mass capture rate variations that wind models predict. 
The centrifugal transition model described above with $\chi = 35^{\rm o}$ and 
$\xi = 1$, was then used to calculate the line shown in Fig. \ref{fig2}; this reproduces 
reasonably well the overall behaviour of the light curve. 
Accordingly a luminosity variation of 
a factor of $\gsim 250$ is produced in response to a factor of 
$\sim 2$ variation in $\mdot$: clearly the centrifugal transition 
works like a very efficient amplifier. Moreover, these values of 
$\chi$ and $\xi$ are among those that match nicely the pulsed fraction
versus luminosity variation (see Fig. \ref{fig2}).
By adopting the same model for the equatorial wind of the Be star over the 
phase interval of the August 2000 observation (0.42--0.51), the predicted  
$\mdot$ is a factor of $\sim 10$ smaller (a value for which the neutron star 
is well in the propeller regime) giving rise to a luminosity in good
agreement with the observed value.

\section{Conclusions}

The BeppoSAX observations of the transient X--ray pulsar 4U~0115+63 
revealed for the first time the presence of the extreme variations as the
source approached periastron. 
The luminosities encompassed by these variations are within the range
predicted by modelling of the centrifugal transition, which separates  
the propeller from the direct accretion regime over a small interval 
of mass inflow rates. 
On the contrary during a quiescent state observation the source was most 
probably in the propeller regime, as its luminosity was lower still 
and nearly constant (however see below). 

Before the present work the evidence for the onset of the centrifugal barrier 
was based on the sudden steepening of the outburst decay below a luminosity
of $\sim 10^{36}\ergs$ in V0332+53 (Stella et al. 1986) and 4U 0115+63 
(Tamura et al. 1992). In both cases, the source became quickly undetected 
in the relevant (collimator) instruments, such that the transition towards the 
propeller regime could not be seen. It was also proposed that several other 
sources in their quiescent (or low) state host a neutron star 
in the propeller regime: among these are the HXRTs A 0538--66 (Campana 1997; Corbet 
et al. 1997) and A 0535+26 (Negueruela et al. 2000). 
The evidence reported here is far more convincing in that dramatic source flux 
variations of the kind expected in the centrifugal transition regime were   
observed for the first time. These variations occur within the luminosity
interval predicted by models of disk-magnetospheric interaction for the case
of 4U 0115+63, where the neutron star spin and magnetic field, as well as
distance are fairly accurately measured. 
The source flux level and absence of sizeable variability in the
faint state away from periastron is also in agreement with basic
expectations for the regime in which the centrifugal barrier is fully
closed. Yet, owing to poor statistics a thermal like spectrum cannot
be ruled out. The analogy with neutron star soft X--ray transients
suggests that such a spectral component, if present, might be due
to reemission of heat in the inner crust caused by pycnonuclear reactions 
(Brown, Bilsten \& Rutledge 1998; Campana et al. 1998; Colpi et al. 2001).
The inferred blackbody luminosity\footnote{Note 
that the spectrum emerging from a hydrogen atmosphere would be 
different from a pure blackbody; yet bolometric corrections are relatively 
small (factor of 2--3 depending on the spectrum) and are ignored here.} 
is in $\sim 10^{33}\ergs$ to be compared with an expected deep crustal heating 
luminosity of $\sim 5\times 10^{33}\ergs$ under the hypothesis that 4U~0115+63 
has accreted at a time average rate of $\sim 5\times 10^{15}\gs$ (as suggested 
by the RossiXTE ASM light curves, {\tt http://xte.mit.edu/lcextrct/}) 
for some $\sim 10^4$ years.
Independent of the origin of the quiescent emission from
4U~0115+63, the data presented in this paper provide substantial
new evidence in favor of the centrifugal transition regime.

\acknowledgments
This research has made use of SAXDAS linearized and cleaned event
files (Rev.2.0) produced at the BeppoSAX Science Data Center.
This work was partially supported by ASI, Co-fin and CNAA grants.

\end{document}